\documentclass[aps, prb, twocolumn, superscriptaddress, floatfix]{revtex4-1}
\usepackage{amsmath}
\usepackage{amsfonts}
\usepackage{braket}
\usepackage[hidelinks]{hyperref}
\usepackage{graphicx}
\usepackage{subfigure}
\graphicspath{ {C:/Users/Arman/Documents/Research/Hybrid Qubit} }

\newcommand{\UMBC}{Department of Physics, University of Maryland Baltimore County, Baltimore, Maryland 21250, USA}

\begin{document}

\title{Rapid adiabatic gating for capacitively coupled quantum dot hybrid qubits without barrier control}

\author{A.~A.~Setser}
\affiliation{\UMBC}
\author{J.~P.~Kestner}
\affiliation{\UMBC}

\begin{abstract}
We theoretically examine the capacitive coupling between two quantum dot hybrid qubits, each consisting of three electrons in a double quantum dot, as a function of the energy detuning of the double dot potentials. We show that a shaped detuning pulse can produce a two-qubit maximally entangling operation in $\sim 50$ns without having to simultaneously change tunnel couplings. Simulations of the entangling operation in the presence of experimentally realistic charge noise yield two-qubit fidelities over $90$\%. 
\end{abstract}

\maketitle

\section{Introduction}\label{intro}
Spin qubits in semiconductor quantum dots are attractive building blocks for quantum computers due to their small size and potential scalability. Single-spin qubits are a simple design in which single-electron spin states are used as the logical basis for computation\cite{LossPRA1998}. These spin qubits have been realized experimentally in both one-qubit and two-qubit exchange-coupled settings\cite{MorelloNature2012, DzurakNature2015, VandersypenNature2018}. Singlet-triplet qubits are another common type of spin qubit, where the logical basis is formed by singlet and triplet spin states\cite{PettaScience2005, BluhmNature2010}. Capacitive coupling is an attractive choice for two-qubit operations in singlet-triplet systems, due to the relatively simple experimental implementation and lack of leakage. Two-qubit entangling gates in these systems have been discussed theoretically\cite{StepanenkoPRB2007, RamonPRB2010, RamonPRB2011, FernandoRWA2018} and recently demonstrated experimentally\cite{NicholNature2017, ShulmanScience2012}.
Typical gate times for capacitively coupled singlet-triplet qubits are on the order of hundreds of nanoseconds, making them generally susceptible to low-frequency charge noise unless special measures are taken\cite{RalphARXIV}. 
A more recent type of spin qubit is the so-called hybrid qubit, which is encoded in the total spin state of three electrons in a double quantum dot, which allows for fully electrical control \cite{coppersmithPRL}. In that setting, capacitively coupled two-qubit gates are predicted to be shorter than typical entangling gates for singlet-triplet qubits\cite{MehlARXIV}.

In this paper, we examine adiabatic gates between strictly capacitively coupled hybrid qubits within the two-qubit logical subspace. This is a different situation than in Refs.~\onlinecite{MehlPRB2015,MehlARXIV}, which permitted tunneling between qubits and considered diabatic gates. By setting the exchange interactions between qubits to zero in our case, the number of possible leakage states is reduced, typically leading to leakage errors significantly smaller than in Refs. \onlinecite{MehlARXIV, MehlPRB2015}. The charge-like character of the hybrid qubit at small detunings gives rise to a large coupling strength while the spin-like character at large detunings effectively turns off the interaction between qubits, and recently Ref.~\onlinecite{wisconsin} has shown that adiabatic pulses in the detuning can be used to perform entangling gates.  Our work differs from Ref.~\onlinecite{wisconsin} in two ways: i) Ref.~\onlinecite{wisconsin} considers simple sinusoidal ramp shapes, whereas we allow for shaped pulses of the detuning, and ii) Ref.~\onlinecite{wisconsin} allows different detunings for each qubit and/or simultaneous control over the tunnel couplings, whereas we restrict to only symmetric detuning control.

By choosing an optimal pulse shape for the detuning, we show that two-qubit entangling operations can be performed in under $50\text{ns}$ while maintaining adiabaticity. We then show that these short gate times give rise to qubits which are naturally robust against realistic charge noise, giving fidelities over $90\%$. This performance is comparable to the results obtained by Ref.~\onlinecite{wisconsin} with an alternate approach.

\section{Model}\label{model}
A single hybrid qubit consists of three electrons in a double quantum dot (DQD). For a system of two hybrid qubits, each DQD confines the three electrons in the lowest two valley-states of each dot. The first and second qubits are respectively centered at a positions $\pm R$ with respect to the origin, giving a total separation of $2R$. Each quantum well of a single DQD is centered at $\pm a$ with respect to the center of the DQD.

We consider states with spin $S=1/2$ and $S_z=-1/2$. The possible spin states can be written as $\ket{\cdot S}, \ket{\cdot T}, \ket{S \cdot}$ and $\ket{T\cdot}$, where $\ket{\cdot S}=\ket{\downarrow}\ket{S}, \ket{\cdot T}=\sqrt{\frac{1}{3}}\ket{\downarrow}\ket{T_0}-\sqrt{\frac{2}{3}}\ket{\uparrow}\ket{T_-}, \ket{S\cdot}=\ket{S}\ket{\downarrow}$ and $\ket{T\cdot}=\sqrt{\frac{1}{3}}\ket{T_0}\ket{\downarrow}-\sqrt{\frac{2}{3}}\ket{T_-}\ket{\uparrow}$. The singlet, unpolarized triplet, and polarized triplet states are respectively represented by $\ket{S}, \ket{T_0}$, and $\ket{T_-}$. In this notation, $\ket{\cdot S}$ and $\ket{\cdot T}$ lie in a $(1, 2)$ configuration, while $\ket{S\cdot}$ and $\ket{T\cdot}$ lie in a $(2, 1)$ configuration. Depending on which direction the double well is biased, either $\ket{\cdot T}$ or $\ket{T\cdot}$ is a high energy state and can be neglected. In the basis of the remaining low energy states (either $\left\{\ket{\cdot S}, \ket{\cdot T}, \ket{S\cdot}\right\}$ or $\left\{\ket{S \cdot},\ket{T\cdot}, \ket{\cdot S} \right\}$), the Hamiltonian for the $i$th qubit is given by \cite{KohPRL2012},
\begin{equation}\label{H1}
H_i=\begin{pmatrix}
-\frac{\varepsilon_i}{2} & 0 & \Delta_1^{(i)} \\
0 & -\frac{\varepsilon_i}{2}+E_{ST}^{(i)} & \Delta_2^{(i)} \\
\Delta_1^{(i)} & \Delta_2^{(i)} & \frac{\varepsilon_i}{2}
\end{pmatrix}, \end{equation}
where $\varepsilon_i$ is the detuning (i.e., the energy difference between the two wells) of the $i$th qubit, and $E_{ST}^{(i)}$ is the singlet-triplet energy splitting of two-spin states in a single well of the $i$th qubit. Given that typical fitted values for $E_{ST}$ are significantly smaller than the orbital splitting in this reduced Hilbert space approximation\cite{friesenARXIV}, we assume that the singlet-triplet spin states of a single well occupy different valleys, rather than different orbitals. Assuming the confining potential is parabolic around the minimum of each well, the lowest two electronic wavefunctions can then be approximated by the valley-state wavefunctions given in Ref.~\onlinecite{CulcerPRB2010}. 
$\Delta_1^{(i)}$ represents the $\ket{\cdot S}\leftrightarrow\ket{S\cdot}$ transition for the $i$th qubit and $\Delta_2^{(i)}$ represents the  $\ket{\cdot T}\leftrightarrow\ket{S\cdot}$ or $\ket{T\cdot }\leftrightarrow\ket{\cdot S}$ transition for the $i$th qubit. A $\ket{\cdot S}\leftrightarrow\ket{\cdot T}$ or $\ket{S\cdot}\leftrightarrow\ket{T\cdot}$ transition is not allowed, since these states occupy different valleys.

Assuming the barrier between qubits is high enough that interqubit tunneling is negligible and the interaction between qubits is purely capacitive, we can incorporate the two-qubit interaction through a set of two-electron Coulomb integrals. There are three terms, corresponding to the three possible types of overall charge configurations: $(2, 1, 1, 2)$, $(1, 2, 1, 2)$/$(2, 1, 2, 1)$, or $(1, 2, 2, 1)$, where $(i, j, k, l)$ denotes the charge configuration $(i, j)$ of the first qubit and $(k, l)$ of the second qubit. The spatial distribution of these three charge configurations leads to three unique Coulomb interactions, which provides a nonzero energy difference between two-qubit states in separate charge configurations. The charge configuration of a low-energy eigenstate is a detuning-dependent mixture of the three types of overall charge configurations, and hence the detuning provides a means of controlling the inter-qubit interaction energy.

We let $V_f, V_m, V_n$ denote the Coulomb interaction between $(2, 1, 1, 2)$, $(2, 1, 2, 1)$, and $(1, 2, 2, 1)$ states, i.e., \textit{far}, \textit{medium}, and \textit{near} interactions. Since the direct Coulomb integral between valley state wavefunctions centered at positions $\bf{r}_1$ and $\bf{r}_2$ simplifies to a direct Coulomb integral between ground-state harmonic wavefunctions centered at $\bf{r}_1$ and $\bf{r}_2$, the interaction terms are given by,
\begin{align}
V_f&=\langle \phi_{-R-a}\phi_{+R+a}|\mathcal{C}|\phi_{-R-a}\phi_{+R+a}\rangle, \notag \\
V_m&=\langle \phi_{-R-a}\phi_{+R-a}|\mathcal{C}|\phi_{-R-a}\phi_{+R-a}\rangle, \label{coulomb} \\
V_n&=\langle \phi_{-R+a}\phi_{+R-a}|\mathcal{C}|\phi_{-R+a}\phi_{+R-a}\rangle ,\notag \\ \notag
\end{align}
where the general integral, $\langle\phi_{R_i}\phi_{R_k} |\mathcal{C}|\phi_{R_j}\phi_{R_l}\rangle$, is presented in Appendix \ref{integrals}.

Tuning the quantum dots so that the low energy basis states of the first and second qubits are respectively given by $\{\ket{\cdot S}, \ket{\cdot T}, \ket{S\cdot}\}$ and $\{\ket{S \cdot},\ket{T\cdot}, \ket{\cdot S} \}$ (i.e., raising the energy of the left well of the left qubit and the right well of the right qubit), allows for a majority of the first and second qubit's states to lie respectively in the  $(1, 2)$ and $(2, 1)$ charge configurations. This gives the largest number of \textit{near} interactions, and hence the strongest coupling between qubits. Summing the single-qubit Hamiltonians and including an interaction term, the two-qubit Hamiltonian is given as, 
\begin{equation}\label{H2}
H=H_1 \otimes I+I\otimes H_2+H_{\text{int}}.
\end{equation}
Assuming the two-qubit basis is a Kronecker product of the single-qubit basis, the interaction Hamiltonian from the direct Coulomb coupling is given by $H_{\text{int}}=\text{diag}(V_n, V_n, V_m, V_n, V_n, V_n, V_m, V_m, V_f)$ (see Appendix \ref{Hamiltonian}).

\section{Effective Hamiltonian}\label{effective}
We form the effective Hamiltonian by restricting the evolution to the four lowest energy states. The effective Hamiltonian in this subspace can be written in the basis of detuning-dependent instantaneous eigenstates as $H_{\text{eff}}=\text{diag}(E_1, E_2, E_3, E_4)$, where $E_n$ is the $n$th smallest eigenvalue of the full Hamiltonian. 
In terms of the $SU(4)$ generators, we can also write
\begin{equation}\label{Heff1}
H_{\text{eff}}=J_{ZI}\sigma_{ZI}+J_{ZZ}\sigma_{ZZ}+J_{IZ}\sigma_{IZ},
\end{equation}
up to a constant term, where $\sigma_{ij}\equiv\sigma_i\otimes \sigma_j$, $J_{IZ}=1/4\left(E_1-E_2+E_3-E_4 \right)$, $J_{ZI}=1/4\left(E_1+E_2-E_3-E_4 \right)$, and $J_{ZZ}=1/4\left(E_1-E_2-E_3+E_4\right)$. As long as the detuning is changed adiabatically, no transitions between adiabatic eigenstates are induced.  Local rotations about the $Z$-axis are induced by $J_{ZI}$ and $J_{IZ}$, while $J_{ZZ}$ generates entanglement. Analytical expressions in terms of the Schrieffer-Wolff approximation are sometimes useful, but we are interested in the small-detuning regime where $J_{ZZ}$ is large, and a perturbative form of $H_\text{eff}$ is not valid; we therefore simply diagonalize the full $9\times 9$ Hamiltonian numerically.

Matching typical silicon-based single-qubit experiments for hybrid qubits,
we take an effective electron mass of $0.2m_0$ ($m_0$ is the electron rest mass), a dielectric constant of $\kappa=11.7\epsilon_0$, a confinement energy of $\hbar\omega=0.38\text{meV}$ (giving a Bohr radius of roughly $31\text{nm}$), and an intraqubit distance of $2a=135\text{nm }$\cite{friesenARXIV}. We choose energy splittings and tunnel couplings of $E_{ST}^{(1)}=52\mu\text{eV }$, $E_{ST}^{(2)}=47\mu\text{eV }$, $\Delta_1^{(i)}=0.64 \times E_{ST}^{(i)}$, and $\Delta_2^{(i)}=0.58\times E_{ST }^{(i)}$,  which minimizes the effect of charge noise on the single qubit terms, $J_{IZ}$ and $J_{ZI}$\cite{wisconsin}. The interqubit distance is taken arbitrarily to be $2R=8a\approx543\text{nm}$, which is similar in scale to non-capacitively coupled two-qubit silicon devices\cite{DzurakNature2015}. The Coulomb interaction terms are then $V_f=181\mu\text{eV}, V_m=227\mu\text{eV},$ and $V_n=303\mu\text{eV}$.

\begin{figure}[t]
{\includegraphics[width=\columnwidth]{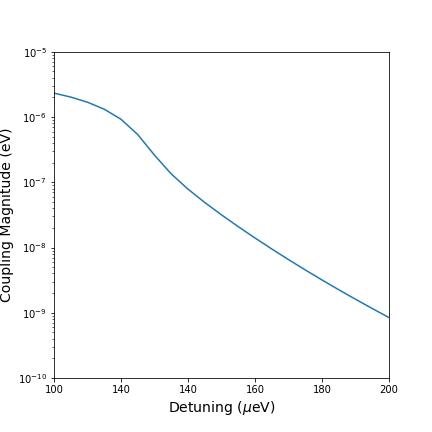}}
\caption{Semi-log plot of the magnitude of the coupling term, $J_{ZZ}$, with $\varepsilon_1=\varepsilon_2=\varepsilon$, as a function of detuning.}
\label{ZZ plot}
\end{figure}

It should be noted that the $9\times 9$ Hamiltonian implicitly assumes that only the lowest orbital can be populated. 
In general, there may be orbital excitations as well. Matrix elements which couple the $9\times 9$ Hamiltonian to these higher energy terms can be shifted into the $9\times 9$ Hamiltonian and treated perturbatively using the Schreiffer-Wolff transformation\cite{winkler, LossSW, coppersmithPRL}. The $n$th order perturbation term will go like $t^{n+1}/(\Delta U)^n$, where $t$ is the transition rate to the higher-energy states and $\Delta U$ is the energy gap between the high-energy states and low-energy states. Since the transition rate is related to the movement of an electron into an excited orbital within a single well, it is approximated by the Coulomb integral $\langle \phi_{R_i}\phi_{R_j} |\mathcal{C}|\phi_{R_k}\widetilde{\phi}_{R_l}\rangle\approx 0.1\mu\text{eV}$, where $\widetilde{\phi}_{R_m}$ denotes an orbital excitation centered at $R_m$, and $R_i, R_j, R_k$ and $R_l$ are assigned the same numeric value (see Appendix \ref{integrals}). Assuming $\Delta U\sim\hbar\omega=0.38$meV, the largest of the perturbative terms will be approximately $50 \text{ peV}$, which is more than an order of magnitude smaller than the minimum value of $J_{ZZ}$ we consider (see Figure \ref{ZZ plot}). Under these assumptions, the $9\times 9$ Hamiltonian accurately approximates the total Hilbert space.

Figure \ref{ZZ plot} shows the effect of the detuning on the effective coupling strength, where we set $\varepsilon_1=\varepsilon_2=\varepsilon$ for simplicity. At small detunings, the eigenstates of the effective Hamiltonian contain mixtures of the three unique charge states, with the resulting state-dependent charge density leading to a large interaction.
Increasing the detuning causes all eigenstates of the effective Hamiltonian to remain approximately in a $(1, 2, 2, 1)$ charge configuration, effectively turning off the state-dependent interaction.

\section{Adiabatic Ramp}\label{ramp}

Both qubits are typically parked at an idle position at large detuning where the interaction is negligible, which we denote by $\varepsilon_{\text{init}}$.  The two logical states of each qubit are defined as the lowest two eigenstates at that detuning. In order to perform an entangling operation, we adiabatically lower the detuning over a time $t_{\text{ramp}}$ to a strongly interacting detuning $\varepsilon_{\text{wait}}$ where the qubits are held for a time $t_{\text{wait}}$. The detuning is then adiabatically returned back to $\varepsilon_{\text{init}}$. Thus, at the end of the pulse, minimal population has been transferred, and the qubits have picked up a nonlocal state-dependent phase. 

\begin{figure}[t]
{\includegraphics[width=\columnwidth]{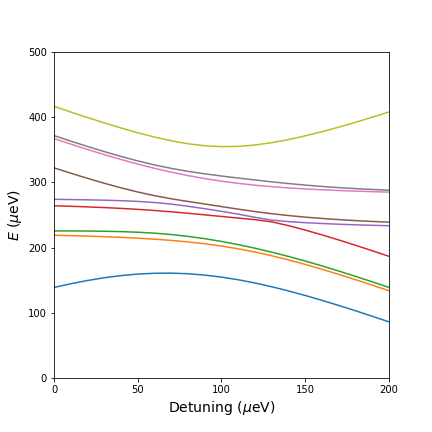}}
\caption{The spectrum of the Hamiltonian assuming $\varepsilon_1=\varepsilon_2=\varepsilon$. At large detunings, the four lowest energy levels (the logical subspace) are approximately parallel, signifying a coupling close to zero. As the detuning decreases, the logical subspace approaches the leakage space, causing an increased interaction}
\label{spectrum}
\end{figure}

We set $\varepsilon_{\text{init}}=200\mu\text{eV}$, so that the coupling is approximately $0$ at the beginning and end of the ramp. As seen in Figure \ref{spectrum}, an avoided crossing between the logical subspace and leakage space occurs roughly around $\varepsilon=130\mu\text{eV}$. Choosing a value of $\varepsilon_{\text{wait}}$ below this point will require a long ramp time in order for the adiabatic approximation to be satisfied. For this reason, we restrict ourselves to $\varepsilon_{\text{wait}}\geq130\mu\text{eV}$.

\begin{figure}[t]
{\includegraphics[width=\columnwidth]{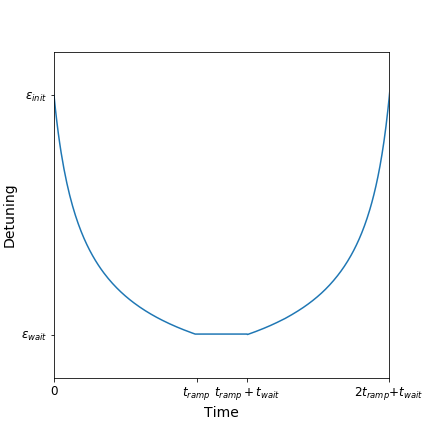}}
\caption{An example of the total pulse. The detuning is lowered from $\varepsilon_{\text{init}}$ to $\varepsilon_{\text{wait}}$ over a time $t_{\text{ramp}}$. It is held at this point for a time $t_{\text{wait}}$, before being raised back to $\varepsilon_{\text{init}}$ over a time $t_{\text{ramp}}$. Here, $\alpha/\hbar=79.0$, $t_{\text{ramp}}=8.0\text{ns}$, $t_{\text{wait}}=2.9\text{ns}$, $\varepsilon_{\text{init}}=200\mu\text{eV}$, and $\varepsilon_{\text{wait}}=145\mu\text{eV}$.}
\label{pulse shape}
\end{figure}

Given that the coupling increases quickly as the detuning approaches the avoided crossing, it is useful to choose a pulse such that $\dot{\varepsilon}$ decreases as $\varepsilon\rightarrow\varepsilon_{\text{wait}}$. This ensures that the detuning will vary quickly when the gap between the logical and leakage space is large, and will vary slowly as the gap shrinks, minimizing nonadiabatic population loss into the leakage space. Such a pulse can be found as
the numerical solution to the differential equation,
\begin{equation}\label{pulse}
\dot{\varepsilon}(t)=\frac{1}{\alpha}\left(\Delta E(\varepsilon)\right)^2, \quad \varepsilon(0)=\varepsilon_{\text{init}}, \quad t\in\left[0,t_{\text{ramp}}\right]
\end{equation}
where $\Delta E(\varepsilon)$ is the detuning-dependent energy difference between the fourth and fifth adiabatic eigenstates, $\alpha$ is an arbitrary scaling factor which allows for control over the ramp time, and $t_{\text{ramp}}$ is defined via $\varepsilon(t_{\text{ramp}})=\varepsilon_{\text{wait}}$\cite{RichermePRA2013}.  The detuning is swept back to its initial value via the time-reversed ramp shape. An example pulse shape is shown in Figure \ref{pulse shape}.

In the noise-free, adiabatic approximation, the ideal evolution operator is 
\begin{equation}
    \widetilde{U}_{\theta}=\exp\left[-i\int_0^{t_{\text{wait}}+2t_{\text{ramp}}} dt H_{\text{eff}}/\hbar\right],
\end{equation}
where, for a given ramp time, the wait time is chosen such that the nonlocal phase acquired over the duration of the pulse is the desired angle, $\theta = \int_0^{t_{\text{wait}}+2t_{\text{ramp}}} dt J_{ZZ}/\hbar$. For a realistic simulation of the two-qubit operations, we can also consider the effects of noise on the qubits. The effects of charge noise on the qubits are modeled by random static perturbations in the detuning drawn from a Gaussian distribution with an experimentally measured standard deviation of $\sigma=4.4\mu\text{eV}$\cite{erikssonNPJ}, i.e., $\varepsilon_{i}(t)\rightarrow\varepsilon(t)+\delta\varepsilon_{i}$, with $\delta\varepsilon_{i}$ independent of time and unique for each qubit. In addition, finite ramp times contribute nonadiabatic leakage.  Thus, to obtain the actual evolution, when targeting a nonlocal phase $\theta$, we numerically solve Schrodinger's equation for the full $9\times 9$ Hamiltonian, using the ``odeint" package available in SciPy\cite{scipy}. This gives the full evolution operator, which includes the effects of charge noise as well as leakage. We then project the full evolution operator onto the lowest four eigenstates of the full Hamiltonian at $\varepsilon=\varepsilon_{\text{init}}$ (i.e., the logical basis) to get the effective (nonunitary) evolution operator, $U_{\theta}$. 

To target a maximally entangling operation, we choose $\theta=\pi/4$ so that the operation is locally equivalent to $\exp\left[{-i\pi\sigma_{ZZ}/4}\right]$. Note that this is sufficient, along with local rotations, to form a universal gate set. The fidelity between the noisy and noise-free evolution operators, $F(U_{\pi/4}, \widetilde{U}_{\pi/4})$, is calculated using the two-qubit fidelity defined in Ref.~\onlinecite{CabreraScience2007},
\begin{equation}\label{fidelity}
F(U_1, U_2)=\frac{1}{16}\left(4+\frac{1}{5}\sum_{i,j\in\left\{I, X, Y, Z\right\}}\text{tr}\left(U_{1}\sigma_{ij}U_{1}^{\dagger}U_{2}\sigma_{ij}U^{\dagger}_{2}\right)\right),
\end{equation}
averaged over $500$ noise realizations.

\begin{figure}[t]
{\includegraphics[width=\columnwidth]{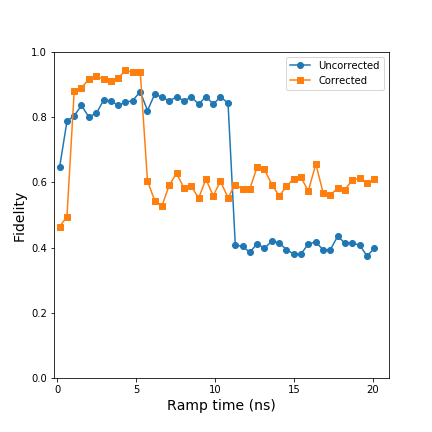}}
\caption{Fidelity of the effective evolution operator versus ramp time, for both the corrected and uncorrected sequences, at $\varepsilon_{\text{wait}}=145\mu\text{eV}$. $t_{\text{wait}}$ is chosen at each point so that the noisy and noise-free operations in the logical subspace are locally equivalent to $U_{\text{target}}$.} 
\label{fidelity plot}
\end{figure}

The choice of $\alpha$ (thus, $t_{\text{ramp}}$) is arbitrary. Increasing the value of $\alpha$ serves to increase the ramp time, thus reducing errors due to leakage. Errors due to charge noise can be suppressed by considering the Hamiltonian in the adiabatic frame, Eq. \ref{Heff1}. The effect of charge noise on the terms in the Hamiltonian can be quantified by $\partial J_{ZZ}/\partial\varepsilon$, $\partial J_{IZ}/\partial\varepsilon$, and $\partial J_{ZI}/\partial\varepsilon$, which are all on the same order of magnitude. Fluctuations in these terms can be suppressed by interweaving applications of specific single-qubit operations in between applications of the noisy two-qubit operation. Specifically, $J_{IZ}$ and $J_{ZI}$ fluctuations can be suppressed completely with the sequence,
\begin{equation}\label{pulse sequence}
U_{\text{corrected}}(\pi/4)=U_{\pi/8}\sigma_{XX}U_{\pi/8}\sigma_{XX},
\end{equation}
where $\sigma_{XX}$ is a local $\pi$ rotation about $\sigma_X$ on both qubits. Assuming essentially instantaneous single-qubit operations relative to the two-qubit gate times and negligible infidelities, we again numerically characterize the full evolution operator as before, except that now the Schrodinger equation is solved for a pulse which is raised and lowered twice, with a nonlocal phase of $\pi/8$ accumulating over each pulse. The fidelity $F(U_{\text{corrected}}, \widetilde{U}_{\pi/4})$ versus ramp time is shown in Fig.~\ref{fidelity plot}. 

Optimizing over $\varepsilon_{\text{wait}}$, we found that the largest fidelity over all values of $t_{\text{ramp}}$ was produced at approximately $\varepsilon_{\text{wait}}=145\mu\text{eV}$, which is the value we use in the plot. For the simple uncorrected operation, we achieve a maximum fidelity of approximately $87.8\%$ at $t_{\text{ramp}}=5.3\text{ns}$. For the corrected operation, we achieve a maximum fidelity of approximately $94.3\%$ at $t_{\text{ramp}}=4.3\text{ns}$. Sub-ns ramp times have low fidelity, due to large adiabatic errors close to $25\%$ and $50\%$ for the uncorrected and corrected operations respectively. Increasing the ramp time quickly lowers errors due to nonadiabaticity below $1\%$, which is negligible compared to the errors due to charge noise. For the uncorrected operation, a sharp drop in fidelity is seen around $10\text{ns}$. At this point, the nonlocal phase acquired by ramping up and immediately down (i.e., $t_{\text{wait}}=0$) is larger than $\pi/4$. Regardless of the choice of $t_{\text{wait}}$, the nonlocal phase acquired by the evolution operator will be larger than $\pi/4$. Since the evolution operator is periodic in $\theta$, we must then choose $t_{\text{wait}}$ so that $\theta=\pi/4+2\pi$, leading to $t_{\text{wait}}$ close to $80\text{ns}$ and hence lower fidelities due to charge noise. For comparison, ramp times under $10\text{ns}$ have wait times of only a few nanoseconds. A similar effect is also seen in the fidelity of the corrected operation.

This is comparable to the performance of Ref.~\onlinecite{wisconsin} which uses a similar model, but a slightly different scheme. Rather than choosing the form of the ramp function to minimize nonadiabatic errors, Ref.~\onlinecite{wisconsin} considers a sine-squared ramp for experimental simplicity and numerically optimizes simultaneous detuning and tunneling pulses, leading to a maximum fidelity around $90\%$.

\section{Conclusion}\label{conclusion}
We have shown that the Coulomb interaction between two hybrid qubits leads to a significant coupling strength within the logical subspace. Adjustment of the individual detunings allows for control over the charge configurations of the individual qubits, and hence the overall coupling strength. We have shown that this controllability allows for fast entangling operations to be performed in less than $50\text{ns}$.

By carefully choosing the detuning pulse shape and using known single-qubit error-correcting sequences, we have shown that fidelities over $90\%$ can be achieved in the presence of realistic charge noise values, without changing tunnel couplings. Further increase in fidelity at the same noise levels would require going through a narrow avoided crossing in the energy eigenstates to access stronger couplings but while maintaining adiabaticity. This suggests the necessity of some sort of shortcut-to-adiabaticity driving protocol,
%comment: I need to add a citation.
with full optimization on the detuning pulse shape simultaneous with tunnel coupling control, similar to the analysis in Ref.~\onlinecite{wisconsin} but with less restriction on the allowed pulse shapes, in order to further improve the fidelity.  Alternatively, increasing the width of the avoided crossing by changing the tunnel couplings would be useful.
%comment: I wonder what parameter controls that?  I assume larger Delta1 would do the trick.

\section*{Acknowledgements}
We thank Adam Frees for useful discussions. This material is based upon work
supported by the National Science Foundation under Grant No. 1620740 and
by the Army Research Office (ARO) under Grant No. W911NF-17-1-0287.

\appendix
\setcounter{section}{0}

\section{Two-Qubit Hamiltonian}\label{Hamiltonian}

For completeness, we present the two-qubit Hamiltonian given by Eq. \ref{H2} in the main text. The full matrix is given by

\begin{equation}
H=
\begin{pmatrix}
E_0 & 0 & \Delta_1^{(2)} & 0 & 0 & 0 & \Delta_1^{(1)} & 0 & 0 \\
0 & E_1 & \Delta_2^{(2)} & 0 & 0 & 0 & 0 & \Delta_1^{(1)} & 0 \\
\Delta_1^{(2)} & \Delta_2^{(2)} & E_2 & 0 & 0 & 0 & 0 & 0 & \Delta_1^{(1)} \\
0 & 0 & 0 & E_3 & 0 & \Delta_1^{(2)} & \Delta_2^{(1)} & 0 & 0 \\
0 & 0 & 0 & 0 & E_4 & \Delta_2^{(2)} & 0 & \Delta_2^{(1)} & 0 \\
0 & 0 & 0 & \Delta_1^{(2)} & \Delta_2^{(2)} & E_5 & 0 & 0 & \Delta_2^{(1)} \\
\Delta_1^{(1)} & 0 & 0 & \Delta_2^{(1)} & 0 & 0 & E_6 & 0 & \Delta_1^{(2)} \\ 
0 & \Delta_1^{(1)} & 0 & 0 & \Delta_2^{(1)} & 0 & 0 & E_7 & \Delta_2^{(2)} \\
0 & 0 & \Delta_1^{(1)} & 0 & 0 & \Delta_2^{(1)} & \Delta_1^{(2)} & \Delta_2^{(2)} & E_8 \\
\end{pmatrix},\end{equation}
where
\begin{align}
E_0&=V_n-\frac{\varepsilon_1}{2}-\frac{\varepsilon_2}{2} \\
E_1&=E_{ST}^{(2)}+V_n-\frac{\varepsilon_1}{2}-\frac{\varepsilon_2}{2} \\
E_2&=V_m-\frac{\varepsilon_1}{2}+\frac{\varepsilon_2}{2} \\
E_3&=E_{ST}^{(1)}+V_n-\frac{\varepsilon_1}{2}-\frac{\varepsilon_2}{2} \\
E_4&=E_{ST}^{(1)}+E_{ST}^{(2)}+V_n-\frac{\varepsilon_1}{2}-\frac{\varepsilon_2}{2} \\
E_5&=E_{ST}^{(1)}+V_m-\frac{\varepsilon_1}{2}+\frac{\varepsilon_2}{2} \\
E_6&=V_m+\frac{\varepsilon_1}{2}-\frac{\varepsilon_2}{2} \\
E_7&=E_{ST}^{(2)}+V_m+\frac{\varepsilon_1}{2}-\frac{\varepsilon_2}{2} \\
E_8&=V_f+\frac{\varepsilon_1}{2}+\frac{\varepsilon_2}{2}
\end{align}

\section{Two-Electron Coulomb Integrals}\label{integrals}

The general two-electron Coulomb integral between harmonic ground-state harmonic wavefunctions is given in Ref. \onlinecite{FernandoPRB2015} as,
\begin{align}
&\langle \phi_{R_i}\phi_{R_k} |\mathcal{C}|\phi_{R_j}\phi_{R_l}\rangle= \notag \\
&\frac{e^2}{4\pi\kappa}\sqrt{\frac{\pi}{2}}\frac{1}{a_B}\exp{\left[-\frac{1}{4a_B^2}\left(\left(R_i-R_j\right)^2+\left(R_k-R_l\right)^2\right)\right]}\notag \\
&\times \exp{\left[-\frac{1}{16a_B^2}\left(R_i+R_j-R_k-R_l\right)^2\right]}\notag \\ 
&\times I_0\left[\frac{1}{16a_B^2}\left(R_i+R_j-R_k-R_l\right)^2\right] \label{C explicit},
\end{align}
where $I_0$ is the zeroth-order modified Bessel function of the first kind, $a_B$ is the effective Bohr radius, $\kappa$ is the effective dielectric constant, and $R_m$ is the distance from the center of the two DQDs to the center of the respective electron's wavefunction.

We are also interested in evaluating terms which involve the interchange of electrons between different orbitals, such as $\langle \phi_{R_i}\phi_{R_k} |\mathcal{C}|\phi_{R_j}\widetilde{\phi}_{R_l}\rangle$, where $\widetilde{\phi}_{R_m}$ denotes an orbital excitation centered at $R_m$. This integral can be evaluated by noting that $\widetilde{\phi}_{R_m}=\sqrt{2}a_B\partial\phi_{R_m}/\partial R_m$. Using this relationship in $\langle \phi_{R_i}\phi_{R_k} |\mathcal{C}|\phi_{R_j}\widetilde{\phi}_{R_l}\rangle$ and noting that the integral is with respect to the spatial coordinates of the wavefunctions, independent of $R_m$, the derivative can be pulled out of the integral, giving, 
\begin{equation}\label{orbital integral}
\langle \phi_{R_i}\phi_{R_k} |\mathcal{C}|\phi_{R_j}\widetilde{\phi}_{R_l}\rangle=\sqrt{2}a_B\frac{\partial}{\partial R_l}\langle \phi_{R_i}\phi_{R_k} |\mathcal{C}|\phi_{R_j}\phi_{R_l}\rangle,
\end{equation}
where the integral on the RHS is given by Eq. \ref{C explicit}.

\bibliography{bibfile}
 
\end{document}